# ENHANCING PROGRAMMING PAIR WORKSHOPS: THE CASE OF TEACHER PRE-PROMPTING

Johan Petersson
Department of Informatics, CERIS
Örebro University
Johan.petersson@oru.se

**Abstract:**

This paper presents a descriptive case study of a student-centered learning activity supported by generative AI. The activity, called the Programming Pair Workshop, focuses on peer learning and student interaction. The presentation describes the workshop format and the rationale behind it, based on an analysis of the assignment guidelines used by the Information Systems students for whom the activity is intended. In addition to the descriptive account of the learning activity, the paper shares experience-based teacher reflections and student perceptions. Key lessons learned highlight that the inclusion of generative AI lowers the threshold for peer-to-peer discussion and emphasizes the importance of clearly communicating the pedagogical rationale behind its implementation. Another interesting lesson is that the introduction of generative AI can also have unintended negative effects on how students perceive a student-centered learning activity.

## I. INTRODUCTION

Although information systems (IS) is far from being the only complex field of education, it is in many ways this very nature that defines it. In IS education, we aim to train future processionals to become technically competent generalists with a deep understanding of context [Leidig and Salmela, 2022]. This focus is particularly important in the more technical and practical areas, such as programming [Leidig et al., 2019]. Rather than becoming experts in a specific programming language or solving hardware-related problems, IS students need to understand generic solution patterns, the concepts behind them and how to tackle new, "unknown" problems [Cummings and Janicki, 2020]. To obtain this, IS education needs to stress the development of generic competencies. That is, competencies such as learning how to learn and collaborative problem solving [Leidig and Salmela, 2022]. Consequently this mean that IS educators benefit from using pedagogical approaches such as student-centered learning, which have shown to develop *both* understanding for course content and generic competencies [Landry et al., 2008]. Calling for student-centered teaching can, however, be easier than implementing it. A major challenge for teachers is to adequately play the role of facilitator of individual learning when the group of students is large [McCabe and O'Connor, 2014]. This means facilitating active learning interactions among students while also engaging in individual support as a teacher, intervening at critical moments when students will benefit the most. Maintaining the flow of interaction is crucial in student-centered learning activities [McCabe and O'Connor, 2014], so it's important to ensure that discussions continue even when students are confused and that the lack of instant teacher access not hinder the learning process (more than is healthy for the motivation to try independently).

In recent years, technological developments have created new opportunities through the increased maturity and availability of generative AI (GenAI). Beyond the power of peer teaching and student-teacher interaction, there is now the possibility of incorporating LLM-based chatbots as an additional component of an active learning situation. That is, to complement the teacher as facilitator and fellow students as discussion partners to engage even more fruitful learning interaction [Lin and Chang, 2023].

This paper presents a case of such an application and the purpose is to describe, motivate, and reflect its design and initial results. The application is called "pair workshop", and the focus of the paper is specifically on how GenAI has impacted its learning value. For the sake of adoptability, the description and motivations are based on excerpts from the actual guidelines used in class. The reflection is based on teachers' experiences and students' course evaluations from the three semesters the GenAI infused workshop version has been used.

## II. THE CASE OF TEACHER PRE-PROMPTING IN PAIR WORKSHOPS

### Pair Workshops for Student Active Learning

Both the introductory and advanced programming courses in the IS program at Örebro University have included learning activities based in student-centered pedagogy for several years. One of these elements is the pair workshop, a small-group interaction activity that is often combined with on-demand lectures, group workshops and project work in the same course [Petersson et al., 2020]. During a pair workshop, 2-3 students work together in the classroom to solve pre-determined problem-solving and discussion tasks. The basis for this learning activity is peer-learning i.e. that students in the same class learn together and from each other [Falchikov, 2001]. A pair workshop typically lasts 2 hours and is part of a flipped classroom setup where students prepare by watching lecture videos and/or reading relevant course literature. In this implementation, a single instructor manages up to 90 students at a time with the support of two teaching assistants (TAs). The instructor is responsible for introducing the workshop and randomly assigning students to groups. Although the instructor may occasionally interrupt a pair workshop with a brief mini-lecture, the primary role of both the instructor and the TAs is to move between groups, interact, and answer questions. To make this practical, the workshop is held in a large, traditional lecture hall with groups seated in every other row to create their own space and allow teachers and TAs to move freely around the room.

A central component in the design of the pair-workshop is the assignment guidelines that groups follow in each session. Such a document outlines the day's tasks and, perhaps more importantly, directs how the students' learning interactions should take e. Some tasks involve direct programming using a computer and a development environment, while others focus on discussing and defining concepts. Because the tasks involve coding, the model uses pair programming, which means that students have rotating but defined roles [Radermacher et al., 2012]. This also means that the group's coding is done by the person who is the "driver" at the moment, rather than by everyone at the same time. The purpose of using explicit role distribution in pair programming is to encourage discussions about problem-solving patterns and to prevent the most confident student from taking over the coding by rotating roles. Because the assignments focus on conceptual understanding (based on code examples), the guidelines often encourage participants to "go around the team" to explain concepts to each other and compare individual interpretations. The point of this approach is to harness the power of peer learning, where different individuals' understandings are tested against each other, taking knowledge to the next level [Clifford, 1999]. In addition, the starting point for a discussion question often comes from a specific passage in the course literature to highlight and remind students of the benefits of using well-developed and relevant sources (based on our experience that the availability of other sources tends to overshadow the course literature in an undesirable way).

After a pair workshop, there is usually a recorded review with solution proposals where an instructor discusses the different tasks and practices think-aloud programming [Arshad, 2009] The foundation for the videos from the pair workshop is the pedagogical idea of an apprentice approach, where a novice learns by imitating a "master's" actions while also developing an understanding of their underlying motives [Kölling and Barnes, 2008]. The intention is that this occurs after the novice has tested their own and their peers' understanding of the same problem, creating even better conditions for learning.

### Inclusion of Teacher Pre-Prompting

The above description of the setup accurately reflects how the pair workshop is implemented in the current curriculum. What has changed over the past three semesters is the significant integration of generative AI, specifically the chatbot ChatGPT, as a pedagogical tool. The GenAI-infused version of our pair workshops was first implemented in the Spring 2022 semester and has since been used by approximately 250 students. The overall motivation behind the new workshop design is to increase student engagement. That is, to maintain a high level of interaction and increase explanatory effectiveness by leveraging the personalized interaction that a GenAI-based chatbot has the potential to provide [Lin and Chang, 2023; Yilmaz and Karaoglan Yilmaz, 2023]. Specifically, the change involves expanding the workshop assignment guidelines to explicitly instruct the use of the GenAI tool in a variety of ways during the workshop. To guide students' processes in the desired direction, these instructions consist of pre-written prompts that students use as initial input when interacting with the tool. These are segments of text that instructors have constructed and tested, which are then copied, pasted, and executed in the tool's interface. Each "pre-prompt", as referred to in this context, is accompanied by instructions on how to use the response as discussion and work material for student groups.

The next section focuses on teacher reflections and student evaluations regarding the various ways we use pre-prompts. The course evaluations however also reveal some overarching student opinions on our implementation: a positive aspect students emphasize is that the use of GenAI tools is perceived as meaningful in education. The knowledge of these tools is viewed as future-oriented expertise. On the negative side, some students instead express concerns that the workshops have become too narrowly focused on "asking GenAI", signaling that teachers are relinquishing their responsibility to interact and to answer questions. They argue that the pedagogical rationale behind the approach is difficult to understand.

### AI-Integrated Workshop Assignments

In order to describe and reflect upon how our AI integration looks like during the pair workshops, three types of tasks from the assignment guidelines used in an advanced programming course (object-oriented programming in C#) are presented below. Common to all three task types is that their purpose is to stimulate a learning discussion among the students. The reflection is based both on our teaching experiences and on course evaluations from three semesters.

The first example of a typical assignment focuses on basic and general conceptual understanding (see Figure 1 below). In this case, the topic of the assignment is the concept pair synchronous/asynchronous in programming contexts. The pre-prompt is shown in the gray box and is followed by instructions on how the tool's response should be processed and discussed by the group of students. In this case, the prompt is a request for the tool to generate a definition that is not related to programming. The goal is to get a tool response that provides a general explanation while avoiding the inclusion of a programming context.

> Now prompt the following:
>
> *Can you explain the difference between something synchronous and something asynchronous to me please. Keep your answer short and give a simple example that is not related to programming.*
>
> - Now discuss the tool's response with each other and add follow-up instructions to the prompt if you feel it is necessary.

Figure 1: Assignment Guideline Extract: Pre-Prompt as a Basis for General Concept Discussion

In earlier versions of the workshop materials (before the introduction of GenAI), the basis for this type of task was instead that students were encouraged to come up with and discuss their own non-programming examples to initiate the conceptual discussion. This instruction was intended, as it is now, to lead to a subsequent progression of knowledge within the topic (i.e., to serve as an appropriate entry task). The materials worked then as well, but often resulted in insecure students not contributing to the discussion, while more confident students (justified or not) took the lead in a somewhat hesitant discussion. With the new starting point in the teacher's pre-prompt, instructors see a clear consequence in that the discussion starts faster and is livelier. It is possible that the shift in focus from individual accounts to evaluating something that exists (i.e., the text generated by the tool) has lowered the threshold for discussion. Another interesting observation is that some students seem to have had a more negative experience with this particular type of task in the new workshop format. In previous course evaluations, these general conceptual understanding tasks were described as valued discussions in which the more knowledgeable students took on the role of explaining to others. When the new starting point of discussions was instead a "response" from the chatbot, some students seems to perceive the as less valuable. Unlike before, some students now expressed that they "could have just asked the teacher" (instead of prompting the tool). Such a critique of the approach is very interesting because it may be about a changed view of the learning situation. That is, despite our intentions, our new approach may seem to have have changed these student experiences from a student-centered, active learning situation to a more passive one. That is, starting with the (presumably) correct "answer" presented in an understandable way leaves less to discuss, thereby diminishing the value of engaging in the conversation at all.

The second type of task (Figure 2) is also about developing deeper conceptual understanding (the example in the figure is on the same topic as above). This time, however, it is based on existing source code from a program that students have previously attempted to run.

> - Now paste in the code for the **Program** class from the **News Viewer1 project**. (The code that make out the class itself is sufficient).
>
> > Now prompt the following:
> >
> > Please explain the concept of synchronous programming using the following code:
> >
> > [your pasted code]
>
> - Read and discuss the response: do you all understand the description/is the explanation consistent with your understanding from the course readings?
> - Add follow-up instructions to the prompt if you feel it is necessary.

Figure 2: Assignment Guideline Extract: Pre-Prompt as a Basis for Code-Based Concept Discussion

The pre-prompt again focuses on generating conceptual explanations, but now based on a specific problem context and the source code that solves it. The second part of the subsequent instruction is an example of something central to this model-the ability to follow up and deepen the prompt at a level that suits the individual group's needs. In this way, the student's attention is focused on exploiting one of the values of generative AI – the potential för tailored exploration.

From the instructor's perspective, this is perceived to have influenced how students engage with their learning interaction. Compared to the previous setup, the questions and discussions that students bring to instructors are now at a more informed level. Since we as teacher are involved at a 'higher entry level,' we have observed that teacher-student interactions increasingly focus on clarifying complex cases and nuances. Just as before, we are 'used' to ensure that students draw their own conclusions, but now this process involves more advanced content rather than fundamental concepts. Although students do not reflect on this difference in the course evaluations (which is understandable, as they cannot compare), the potential of the tool as a tailored interlocutor for follow-up questions is discussed. An important point mentioned is that students now feel it is easier and more acceptable to ask "stupid questions" (since they see them as less stupid and more informed).

In the third and final example of task types (Figure 3), the focus is on analyzing and ultimately writing one's own source code (in this example, the goal is to understand concepts and syntax related to anonymous types and so-called LINQ). For this type of task, the setup is somewhat reversed from the above. Here, students are instructed to first create a shared interpretation and discuss a shared piece of code. Only then does the pre-prompt come into play, which involves allowing the tool to address the same questions that the students themselves are considering. As a final prompt, students are given the task of coding in practice, based on their newly acquired knowledge (albeit at a more advanced level).

```
var namnVingBreddFraga1 = from enPapegoja in papegojLista
                          where enPapegoja.VingBredd > 25
                          select new { enPapegoja.Namn, enPapegoja.VingBredd };

foreach (var ettSvarsObjekt in namnVingBreddFraga1)
{
    Console.WriteLine(ettSvarsObjekt.Namn + " " + ettSvarsObjekt.VingBredd);
}
```

a) Now code the above and then discuss:

- What does the created question do = what answer does it give?
- What is the syntax for creating such a "temporary" type?
- The correct terminology in C# is *anonymous type*: in what way would you say it is anonymous?

> Prompt ChatGPT with the same questions as above. Then paste your code as a basis. Compare your reasoning with the answers you get. Were you on the same track?

b) Now code the same query expressions as above but using LINQ extension methods and lambda expressions.

Figure 3: Assignment Guideline Extract: Pre-Prompt as a Basis for Follow-Up on Student Discussion

Our experience so far suggests that this type of "reverse" pre-prompting serves as an important complement to the types mentioned above. The difference may seem marginal at first glance, but we believe it to be central. Although the threshold for group discussion is lowered when the tool response is used as a starting point (see above), this same circumstance risks eliminating alternative interpretations and useful misconceptions for learning in advance. In addition, this task setup is seen as a better stimulus for students to critically examine the tool's responses.

There is no specific feedback on this type of task in the student evaluations. However, it is clear from the course evaluations that the "training" in using the tool in different ways is highly valued. Students consider it central to learn how to use a type of tool that they expect to be important in their professional lives.

## III. CONCLUSIONS

Although this case presentation primarily addresses an ongoing development, there are experience-based conclusions we believe others can benefit from. One conclusion is that, from our educator's perspective, the introduction of GenAI interaction has the potential to add an easily accessible aspect of leveled/tailored support in active learning situations. That is, to provide a type of personalized and formative feedback that is otherwise difficult to achieve at the individual level in large groups of students. We find that the introduction of this component helps to maintain engagement by keeping student interactions more lively. Furthermore, we understand this tailored support as having aspects of inclusion, enabling more workshop participants to engage in group discussions. We further perceive this to influence students' confidence in asking questions to instructors and TA:s. A related but different experience is that the nature of the student-teacher interaction seems to be more developed, leading to a slightly different role for the teacher.

Our introduction of GenAI and teacher pre-prompting however also shows a need to be explicit about pedagogical motivations. In other words, the pedagogical rationale for including these tools needs to be explained to students when such an approach is introduced. This is important because, in our case, students sometimes expressed concern that the generated interaction was intended to replace peer and teacher communication. Related to this it has also proven crucial to problematize the accuracy of the AI-generated support (i.e., to emphasize a discussion of the need for being source-critical) and to reaffirm the role that educators believe this support plays in the learning process.

## ABOUT THE AUTHOR

Johan Petersson is an Assistant Professor at Örebro University, Department of Informatics, CERIS. His research interests include information security, electronic commerce, and pedagogy in the teaching of information systems. Johan is a member of the Swedish Information Systems Academy (SISA) Educational Committee and the Association for Information Systems (AIS). His teaching experience encompasses all levels of higher education, from undergraduate to doctoral studies, with a particular focus on programming, qualitative research methodology and higher education pedagogy. Together with his subject colleagues, Johan was awarded the SISA Educational Prize in 2023 for his work on student-centered pedagogy in programming education.